**Surrogate testing of linear feedback processes with non-Gaussian innovations**

Radhakrishnan Nagarajan

**Abstract**

Surrogate testing is used widely to determine the nature of the process generating the given empirical sample. In the present study, the usefulness of phase-randomized surrogates, amplitude adjusted Fourier transform (AAFT) and iterated amplitude adjusted Fourier transform (IAAFT) surrogates on statistical inference of linearly correlated noise with non-Gaussian innovations and their static, invertible nonlinear transforms from their empirical samples is discussed. Existing surrogate testing procedures which retain the auto-correlation function in the surrogates may not be appropriate in the presence of non-Gaussian innovations.

**Author for Correspondence**

Radhakrishnan Nagarajan
Center on Aging, UAMS
629 Jack Stephens Drive, Room: 3105
Little Rock, AR 72212, USA
Email: nagarajanradhakrish@uams.edu
Fax: (501) 526 5830

**1. Introduction**

Surrogate testing has been used widely to statistically infer the nature of the process generating the given data [1-5]. In experimental settings, one often has access only to a *single realization* of the dynamical process. This single realization is *assumed* to be representative of the underlying dynamics and shall be termed as *empirical sample* in the subsequent sections. Such an assumption is valid especially for *stationary* processes whose statistical properties are time invariant.

*1.1 Surrogate Algorithms*

In studies, it is common to use a hierarchy of surrogate algorithms in a systematic manner [1-5] in order to determine the process generating the given empirical sample. Essential ingredients of surrogate testing include (1) a *surrogate algorithm* that retains certain statistical properties of the given empirical sample (2) a *null hypothesis*, and (3) a *discriminant statistic* that can discern the empirical sample from their surrogate counterparts when the null hypothesis is violated. Surrogate algorithms commonly used include *random shuffled surrogates*, *phase-randomized surrogates* (FT), *amplitude adjusted Fourier transform* (AAFT) and *iterated amplitude adjusted Fourier transform* (IAAFT) [1, 3]. The null hypothesis addressed by random shuffled surrogates is that the given process is generated by uncorrelated noise. Rejecting the null using a chosen discriminant statistic sensitive to correlations indicates existence of linear or nonlinear correlations in the given empirical sample. A more sophisticated algorithm is phase-randomized surrogates. The null hypothesis addressed by phase-randomized surrogates is that the given data is generated by a linearly correlated noise with Gaussian innovations [1]. Power-spectrum is related to the second order statistics (i.e. linear correlation) by Weiner-Khinchin theorem [6]. Thus retaining the power-spectrum of the given data in the FT surrogates implies retaining the auto-correlation function. However, there is no constraint on retaining either the amplitude distribution or nonlinear/higher order correlations in the surrogates. Rejecting the null using a discriminant statistic sensitive to nonlinearities (static or dynamical) often leads to the conclusion that the given empirical sample exhibits nonlinear correlations. Subsequently, algorithms such as AAFT and IAAFT [1, 3] are used to infer the nature of the nonlinearity. The null hypothesis addressed by AAFT and IAAFT surrogates is that the given empirical sample is generated by a static, invertible, nonlinear transform of linearly correlated noise with Gaussian innovations. Recently, IAAFT surrogates has been claimed to be superior to AAFT surrogates [3] in retaining the power-spectrum. Unlike phase-randomized surrogates, AAFT/IAAFT surrogates retain both the amplitude distribution and the power spectrum to a high degree of precision [3]. Rejecting the null using a discriminant statistic sensitive to dynamical nonlinearities is attributed to the existence of dynamical nonlinearity in the empirical sample. IAAFT often forms a precursor to inferring deterministic chaos [1]. As a comment, it should be noted that deterministic chaos is an *instance* of dynamical nonlinearity and does not encapsulate the entire class of dynamical nonlinearities. Several discriminant statistics have been proposed in the past. In the present study, we use *approximate entropy* (ApEn) [7]. Its ability to discern dynamics among a wide-range of linear and nonlinear processes [7] are reasons for its choice. The parameters (m, r, N) in the approximate entropy ApEn(m, r, N) estimation procedure were chosen as : m = 2; r = 0.2 times the standard deviation of the given data and N = 4096 for all the data sets in the present study. A detailed description can be found elsewhere [7].

*1.2 Statistical Inference*

In classical surrogate testing [1-4], a combination of parametric and non-parametric statistical tests is used to assess statistical significance. Parametric test reject the null if the *sigmas* (S=$\frac{|\boldsymbol{m}^s - \boldsymbol{m}^o|}{\boldsymbol{s}_s}$ > 2) [1], ($\boldsymbol{m}^s, \boldsymbol{s}_s$) represent the mean and the standard deviation of the discriminant statistic obtained on the $n_{surr}$ surrogate realizations and $\boldsymbol{m}^o$ is the discriminant statistics obtained on the empirical sample. The choice of (S > 2) can be attributed to implicit normality assumptions where 95% of the samples lie within two standard deviations. One-sided non-parametric testing has also been proposed [1-3] to reject the null at a given significance level ($\alpha$). In the present study, the p-value is determined by the number of instances where the discriminant statistic estimated on the empirical sample is strictly greater than those estimated on the surrogates. The number of surrogates was fixed at 99 and corresponds to a significance level of ($\alpha = 0.01$) [1-3]. Since the present study, focuses on synthetic data sets whose model is known we use traditional parametric (ttest) and non-paramteric (wilcoxon-ranksum) test at ($\alpha = 0.01$) [8] to directly compare the distribution of the discriminant statistic obtained on the independently generated empirical samples (100 realizations) to those obtained on their surrogate counterparts [4].

The results presented in this study bring out the fine distinction between nonlinearity and non-normality. Non-Gaussianity is achieved by (a) static, invertible nonlinear transform of first-order linear feedback process with Gaussian innovations, (b) first-order linear feedback process with non-Gaussian innovations and (c) static, invertible nonlinear transform of first-order linear feedback process with non-Gaussian innovations. It is shown that phase-randomized surrogates, AAFT and IAAFT surrogates may not be adequate in identifying the process from empirical samples generated by linear feedback process with non-Gaussian innovations and their nonlinear transforms.

## 2. First-order linear feedback process

$$x_n = \boldsymbol{q}\, x_{n-1} + \in_n \quad \text{where } n = 1...N \; \dots\dots\dots\dots\dots\dots\dots\dots\dots\dots\text{.....................} (1)$$

First-order linear feedback process is represented by expression (1), with *innovations* $\in_n$ sampled from an independent and identically distributed (i.i.d) process with zero mean and unit variance.

It is important to note that each sample $x_n$ generated by (1) is a *linear combination* of innovations $\in_n$, whose weights are determined by the *process parameter* $\boldsymbol{q}$. Thus the above expression represents a *linearly correlated feedback process*, also known as *linearly correlated noise*. While classical definition of linearly correlated noise implicitly assumes Gaussianity of $\in_n$, such an assumption need not necessarily be true in general [9]. The correlation decay is governed by the sign and magnitude of the process parameter $\boldsymbol{q}$. It can be shown that the above process is stationary when $|\boldsymbol{q}| < 1$, i.e. $-1 < \boldsymbol{q} < 1$.

In the present study, we focus on non-Gaussian processes generated by:

(a) Static, invertible nonlinear transform of the linearly correlated noise $x_n = \boldsymbol{q}\, x_{n-1} + \in_n$, $y_n = x_n \sqrt{|x_n|}$ with process parameter $\boldsymbol{q} = 0.95$ and Gaussian innovations $\in_n$ sampled from normally distributed i.i.d process with zero mean and unit variance [3]. In the subsequent discussion we shall refer to this process as NAWGN, which stands for *nonlinear transformed additive white Gaussian noise*.

(b) Linearly correlated noise $y_n = \boldsymbol{q}\, y_{n-1} + \in_n$ with process parameter $\boldsymbol{q} = 0.95$ and non-Gaussian innovations $\in_n$ sampled from an *exponentially distributed* i.i.d process with zero mean and unit variance. In the subsequent discussion we shall refer to this process as AWNGN, which stands for *additive white non-Gaussian noise*. It can be shown analytically second-order cumulants namely: mean and variance, are sufficient statistics to describe Gaussian innovations $\in_n$. However, for $\in_n$ sampled from exponential distribution higher order cumulants are necessary in order to completely describe the process. Consider $f(x) = \boldsymbol{l}\, e^{-lx}, x \geq 0$, the $k^{th}$ cumulant $c_k$ is given by the expression $c_k = (k-1)\mathbf{l}^{-k}$ which exhibits an exponential decay with $k$.

(c) Static, invertible nonlinear transform of the linearly correlated noise $x_n = \mathbf{q}\, x_{n-1} + \in_n$, $y_n = x_n \sqrt{|x_n|}$ with process parameter $\mathbf{q} = 0.95$ and non-Gaussian innovations $\in_n$ sampled from exponentially distributed i.i.d process with zero mean and unit variance. In the subsequent discussion we shall refer to this process as NAWNGN, which stands for *nonlinear transformed additive white non-Gaussian noise*.

The choice of the process parameter $\mathbf{q} = 0.95$ and the static-invertible nonlinearity $y_n = x_n \sqrt{|x_n|}$ is encouraged by the seminal report on IAAFT surrogates [3]. While the result of NAWGN is already documented [3], we include it in order to establish the subtle difference between non-Gaussianity brought about by static, invertible nonlinear transform (a) and non-Gaussian innovations $\in_n$ inherent in the processes (b) and (c). Process (c) is generated similar to that of (a), however the innovations $\in_n$ were samples from non-Gaussian distribution. The inclusion of NAWGN also justifies the choice of approximate entropy as a discriminant statistic.

## 3. Results

NAWGN, AWNGN and NAWNGN processes were generated as described above. After discarding the initial transients the length of the data was fixed at (N = 4096) points. Quantile-quantile (QQ) plots are used widely in statistical literature for qualitative assessment of discrepancy in the distribution across samples [8]. QQ plots of similar distributions are represented by diagonal lines, also shown as a reference in the accompanying figures in the present study.

(A) *Phase-randomized Surrogates (FT)*

Phase-randomization procedure by its very nature preserves the power-spectrum and not the amplitude distribution [1], Fig. 1. QQ plots of the NAWGN, AWNGN and NAWNGN process and their corresponding FT surrogates reveal that the distribution of the empirical sample is not preserved in their FT surrogate counterpart, Figs. 1a, 1c and 1d. However, the power spectrum, hence the auto-correlation of the empirical sample is preserved in the surrogate realization, Figs. 1b, 1d and 1f. Classical parametric surrogate testing (99 surrogates) resulted in sigmas (S ~ 36 >> 2; S ~ 55 >> 2 and S ~ 77 >> 2) for the NAWGN, AWNGN and NAWNGN processes respectively. Subsequent non-parametric analysis with 99 surrogates also rejected the null at ($\alpha$ = 1/(99+1) = 0.01) for NAWGN, AWNGN and NAWNGN processes. The rejection of the null in the case of NAWGN and NAWNGN can be attributed to the presence of nonlinear correlations not preserved in the FT surrogates. However, it can be attributed to higher correlation in the case of non-Gaussian innovations which are not preserved in the case of FT surrogates. The difference between the approximate entropy estimates between the empirical sample and their FT surrogates ApEn(si)- ApEn(so), i = 1…99   generated from NAWGN, AWNGN and NAWNGN processes were skewed to the right of zero, Figs. 2a, 2b and 2c respectively. Indicating that ApEn(si) was significantly greater than those ApEn(so) for all the 99 realizations conforming to the results of the non-parametric test. In order to reject the claim that the observed significance is not due to a particular realization of the empirical sample, 100 independent realizations of NAWGN, AWNGN and NAWNGN processes and their FT surrogates were generated. The approximate entropy estimates on the 100 empirical samples and their surrogate realizations showed minimal overlap across the three cases, Figs. 2d, 2e and 2f respectively. Statistical testing using ttest and wilcoxon-ranksum test  ($\alpha$ = 0.01)  indicated that the distribution of ApEn estimates on the independent empirical samples and their FT surrogate counterparts were statistically significant (p-value < 0.01). From the above results it is clear that FT surrogates may not be useful in statistically inferring linear feedback process with non-Gaussian innovations and their nonlinear transforms from their empirical samples.

(B) *Amplitude Adjusted Fourier Transform (AAFT)*

Unlike phase-randomized surrogates, AAFT surrogates have been found to preserve the amplitude distribution and the power-spectrum under certain implicit assumptions [1, 3], Fig. 3. By retaining the distribution, higher order moments of the given empirical sample are preserved in their surrogate counterpart. Subsequently, these algorithms have been used for statistical inference of non-Gaussian processes [1-3]. The power-spectrum and the amplitude distribution of the empirical sample and its corresponding surrogate for NAWGN, AWNGN and NAWNGN processes are shown in Fig. 3. Parametric testing resulted in sigmas (S ~ 0.57 << 2) for NAWGN, (S ~ 30 >> 2) for AWNGN and (S ~ 15 >> 2) for NAWNGN, respectively. Subsequent, non-parametric testing failed to reject the null in the case of NAWGN, whereas the null was rejected at ($\alpha = 1/(99+1) = 0.01$) for AWNGN and NAWNGN. The difference in the approximate entropy estimates between the surrogate realizations and the empirical sample, ApEn(si)- ApEn(so), i = 1...99, was distributed on either side of zero for NAWGN, Fig. 4a. This has to be contrasted to those obtained on AWNGN and NAWNGN, which was skewed to the right of zero, Fig. 4b and 4c. Thus, the null can be rejected for AWNGN and NAWNGN, unlike NAWGN. In order to reject the claim that the observed significance is due to a particular realization of the empirical sample, we generated 100 independent realizations of NAWGN, AWNGN, NAWNGN and their corresponding AAFT surrogates. The approximate entropy estimates on the 100 independent empirical samples and their surrogate realizations showed considerable overlap in the case of NAWGN, Fig. 4d, whereas they showed minimal overlap for AWNGN and NAWNGN, Figs. 4e and 4f. Statistical testing using ttest and wilcoxon-ranksum test ($\alpha = 0.01$) indicated that the distributions were statistically significant (p-value < 0.01) for AWNGN and NAWNGN. As expected the tests failed to reject the null for NAWGN. It is well known that AAFT surrogates are useful in inferring non-Gaussian processes that are generated by static, invertible, nonlinear transform of linear feedback process with Gaussian innovations. However, they may not be useful in inferring linear feedback process with non-Gaussian innovations and their nonlinear transforms from their empirical samples.

(C) *Iterated Amplitude Adjusted Fourier Transform (IAAFT)*

IAAFT surrogates were recently proposed as an improvement to AAFT surrogates. More importantly, IAAFT has been found to minimize the flatness in the power-spectrum [3]. As with AAFT, IAAFT has been used extensively to infer the nature of empirical samples generated from non-Gaussian processes. The power-spectrum and the amplitude distribution of the empirical data and its corresponding surrogate for NAWGN, AWNGN and NAWNGN are shown in Fig. 5. Parametric testing resulted in sigmas ($S \sim 0.96 \ll 2$) for NAWGN process, ($S \sim 11 \gg 2$) for AWNGN and ($S \sim 5 > 2$) for NAWNGN. Subsequent, non-parametric testing failed to reject the null in the case of NAWGN, whereas the null was rejected at ($\alpha = 1/(99+1) = 0.01$) for AWNGN and NAWNGN. The difference in the approximate entropy estimates between the surrogate realization and the empirical sample, ApEn(si)- ApEn(so), i = 1…99, was distributed on either side of zero for NAWGN, Fig. 6a. This has to be contrasted to those obtained on AWNGN and NAWNGN, which were skewed to the right of zero, Figs. 6b and 6c. Thus, the null can be rejected in AWNGN and NAWNGN, unlike NAWGN. In order to reject the claim that the observed significance is due to a particular realization of the empirical sample, we generated 100 independent realizations of NAWGN, AWNGN and NAWNGN and their corresponding surrogates. Approximate entropy estimates on the 100 empirical samples and their surrogate realizations showed considerable overlap in the case of NAWGN, Fig. 6d, whereas they showed minimal overlap for AWNGN and NAWNGN, Figs. 6e and 6f. Statistical testing using ttest and wilcoxon-ranksum test ($\alpha = 0.01$) indicated that the distributions were statistically significant (p-value $< 0.01$) for AWNGN and NAWNGN. As expected the tests failed to reject the null for NAWGN. IAAFT surrogates are useful in inferring non-Gaussian processes that are generated by static, invertible, nonlinear transform of linear feedback process with Gaussian innovations. However, they may not be useful in inferring linear feedback process with non-Gaussian innovations and their nonlinear transforms from their empirical samples.

## 4. Discussion

Surrogate testing has been used extensively in the analysis if experimental data sets generated from complex systems. More importantly, rejecting the null in the case of AAFT and IAAFT surrogates have been attributed to existence of dynamical nonlinearity in the given empirical sample. Such analyses often form precursor to determining deterministic chaos which represents an instance of dynamical nonlinearity. Alternately, conclusions drawn in favor of dynamical nonlinearities based solely on FT, AAFT and IAAFT surrogate analysis is incomplete. In the present study, it clearly shown that rejection of null across the three surrogate algorithms can occur for a simple first order linear feedback of processes with non-Gaussian innovations and their static, invertible nonlinear transforms. For experimental data sets obtained from complex natural systems, the innovations need not necessarily be Gaussian. The results presented indicate that retaining the power-spectrum, hence the second-order correlation may not be sufficient in discerning nonlinear from non-Gaussian signatures. This is not a failure of the existing surrogate algorithms but as inherent limitation. The present study also encourages more judicious interpretation of surrogate testing results obtained on experimental data sets. Since dynamical noise of linear and non-linear feedback systems need not necessarily be Gaussian.

**Acknowledgements**

I would like to thank the reviewers for helpful comments and suggestions. The present study is supported by funds from National Library of Medicine (1R03LM008853-1) and junior faculty grant from American Federation for Aging Research (AFAR).

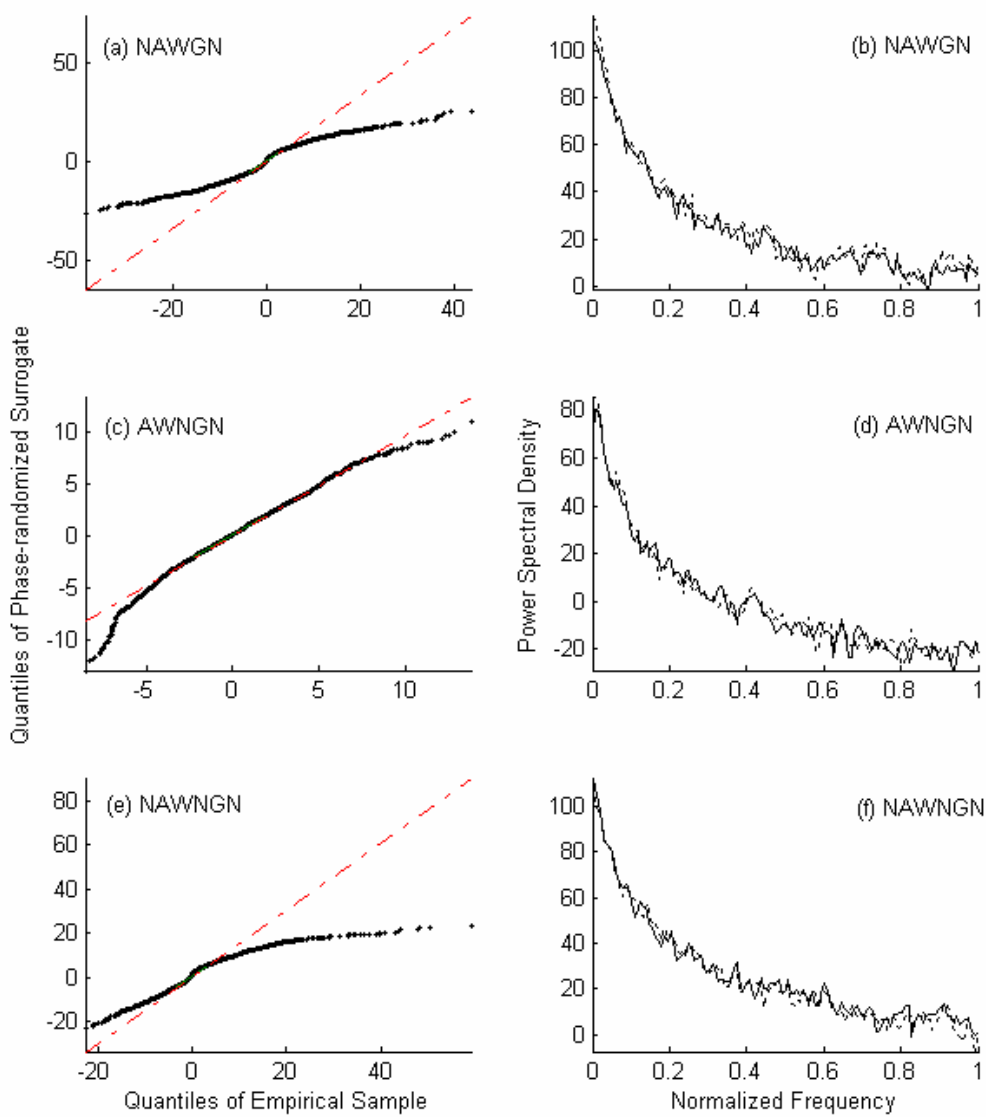

**Figure 1** QQ-plot of the empirical sample and their corresponding FT surrogate counterpart for NAWGN, AWNGN, NAWNGN is shown in (a, c and e) respectively. Their corresponding power-spectra are shown in (b, d and f) respectively. The dashed reference line in (a, c and e) correspond to the case where the distributions of the empirical sample and the FT surrogate are identical. The dotted and solid lines in (b, d and f) correspond to the empirical sample and their FT surrogate counterpart respectively.

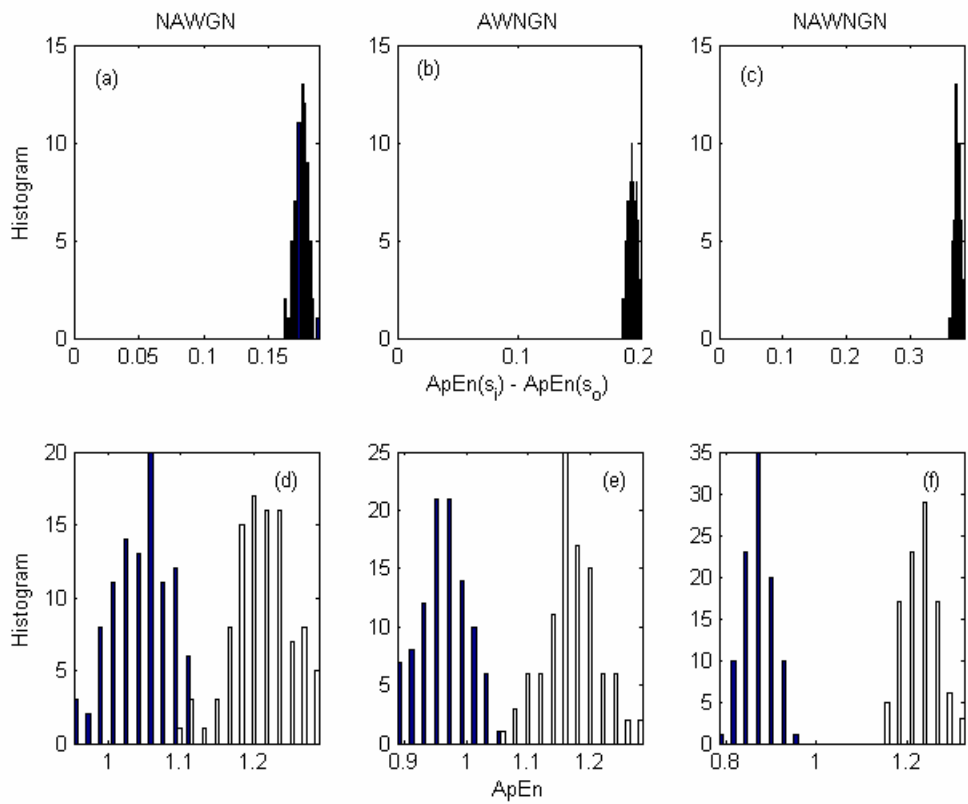

**Figure 2** Distribution of the difference in the approximate entropy estimates between the given empirical sample and their FT surrogate counterparts ApEn($s_i$) – ApEn($s_o$), i = 1…99 for NAWGN, AWNGN and NAWNGN is shown in (a, b and c) respectively. Distribution of the approximate entropy estimates of 100 independent empirical samples (solid bars) and their FT surrogate counterparts (hollow bars) for NAWGN, AWNGN and NAWNGN is shown in (d, e and f) respectively.

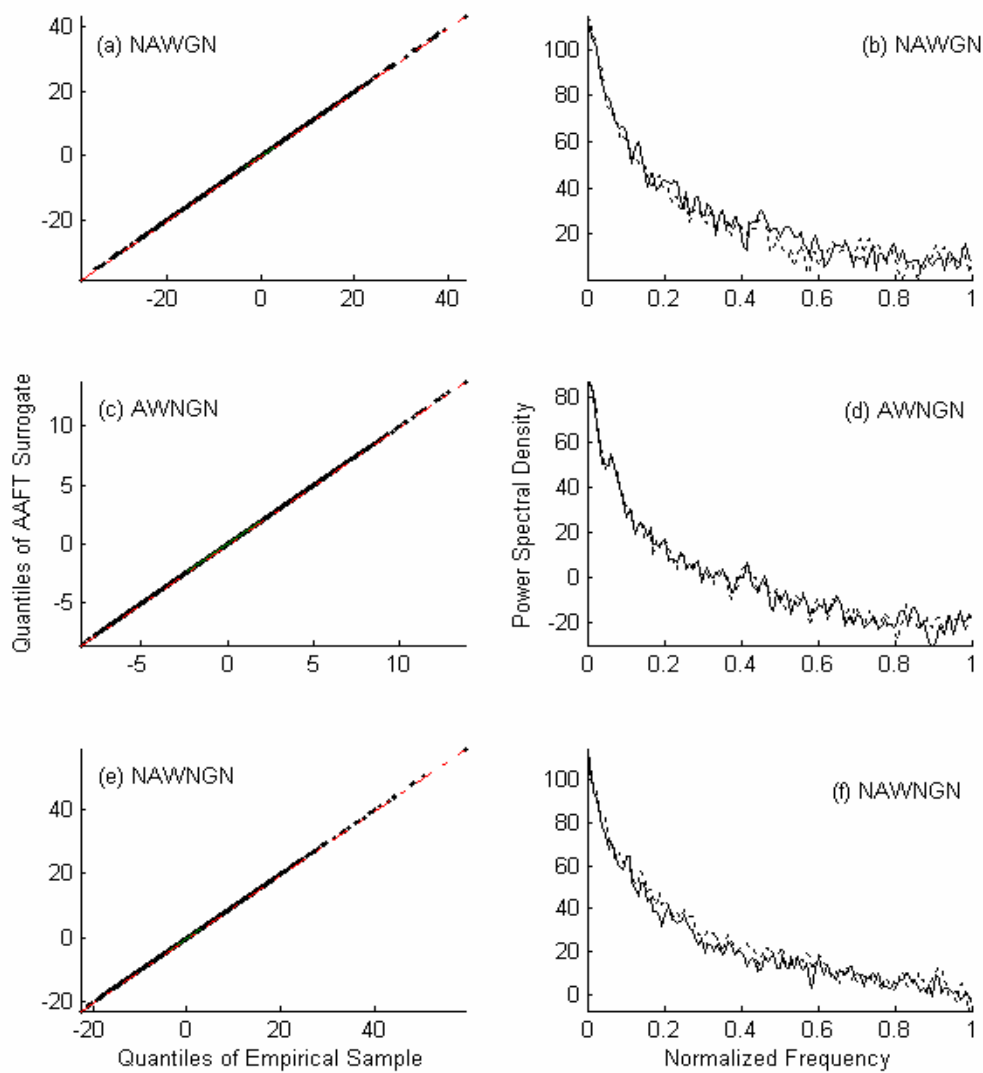

**Figure 3** QQ-plot of the empirical sample and their corresponding AAFT surrogate counterpart for NAWGN, AWNGN, NAWNGN is shown in (a, c and e) respectively. Their corresponding power-spectra are shown in (b, d and f) respectively. The dashed reference line in (a, c and e) correspond to the case where the distributions of the empirical sample and the AAFT surrogate are identical. The dotted and solid lines in (b, d and f) correspond to the empirical sample and their AAFT surrogate counterpart respectively.

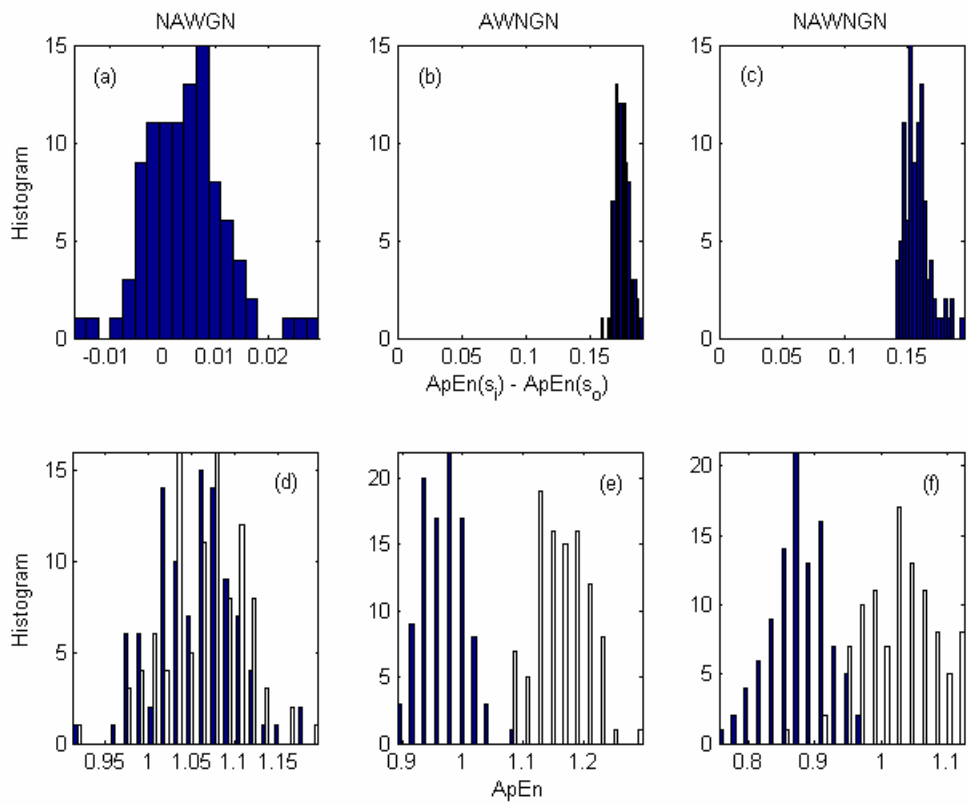

**Figure 4** Distribution of the difference in the approximate entropy estimates between the given empirical sample and their AAFT surrogate counterparts $ApEn(s_i) - ApEn(s_o)$, $i = 1...99$ for NAWGN, AWNGN and NAWNGN is shown in (a, b and c) respectively. Distribution of the approximate entropy estimates of 100 independent empirical samples (solid bars) and their AAFT surrogate counterparts (hollow bars) for NAWGN, AWNGN and NAWNGN is shown in (d, e and f) respectively.

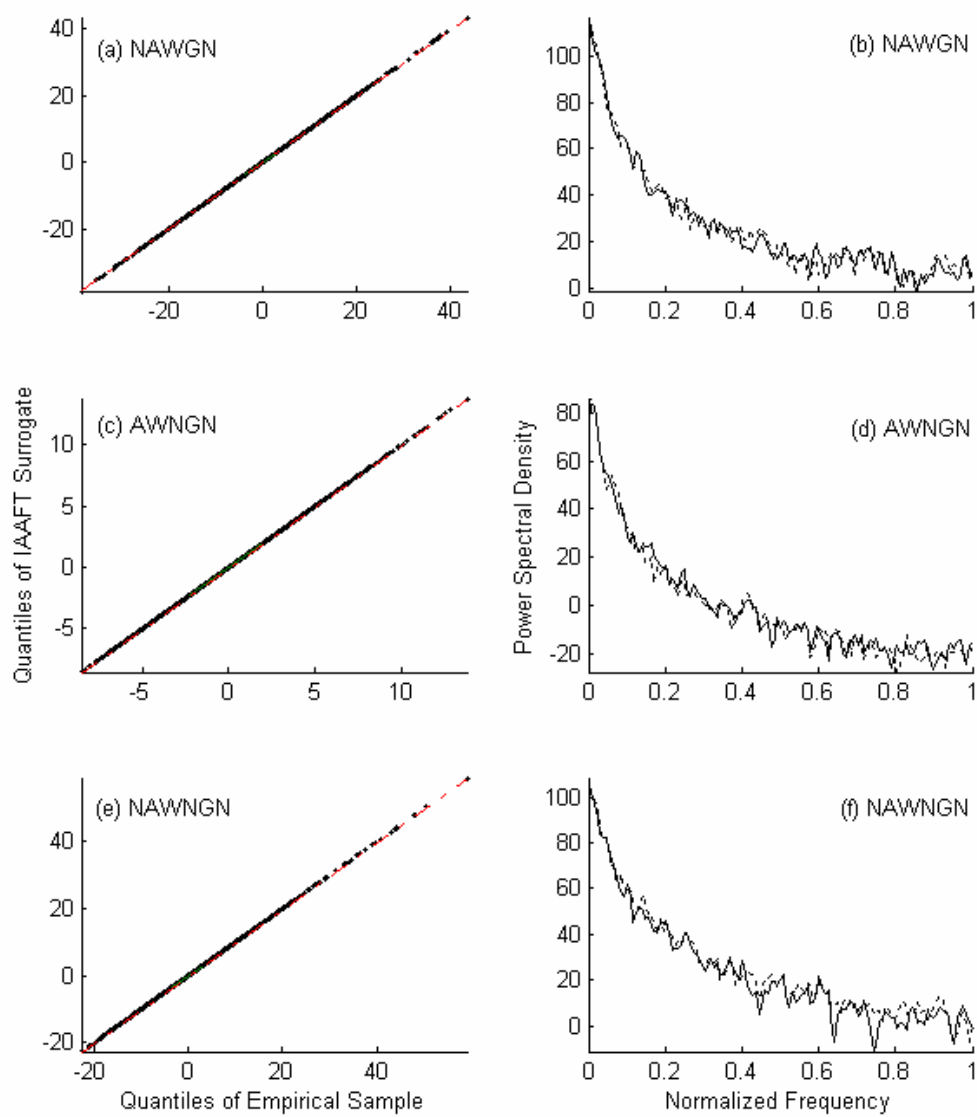

**Figure 5** QQ-plot of the empirical sample and their corresponding IAAFT surrogate counterpart for NAWGN, AWNGN, NAWNGN is shown in (a, c and e) respectively. Their corresponding power-spectra are shown in (b, d and f) respectively. The dashed reference line in (a, c and e) correspond to the case where the distributions of the empirical sample and the IAAFT surrogate are identical. The dotted and solid lines in (b, d and f) correspond to the empirical sample and their IAAFT surrogate counterpart respectively.

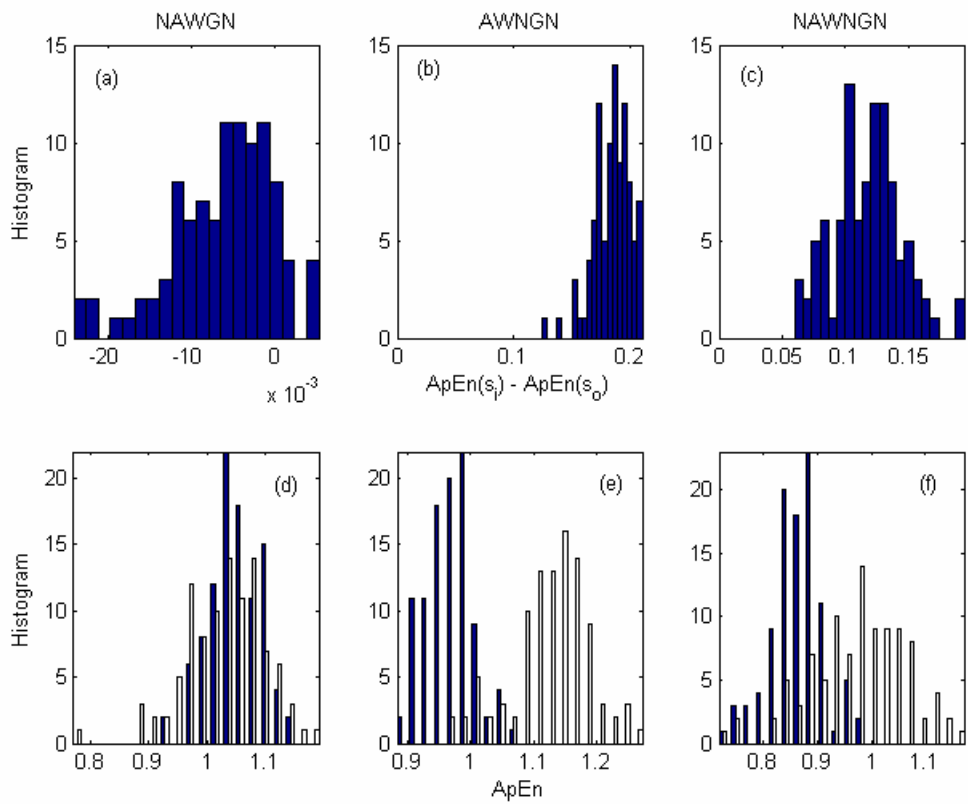

**Figure 6** Distribution of the difference in the approximate entropy estimates between the given empirical sample and their IAAFT surrogate counterparts $ApEn(s_i) - ApEn(s_o)$, i = 1…99 for NAWGN, AWNGN and NAWNGN is shown in (a, b and c) respectively. Distribution of the approximate entropy estimates of 100 independent empirical samples (solid bars) and their IAAFT surrogate counterparts (hollow bars) for NAWGN, AWNGN and NAWNGN is shown in (d, e and f) respectively.

**Appendix I (additional results generated for review not in the original manuscript)**

**Repeating the above analysis using the first-lagged mutual information as the discriminant metric resulted in similar results.**

Mutual information is a generalized measure of dependency and sensitive to higher order correlations, hence its choice. Please see Figure R1 and Table II below. Mutual information was estimated using uniform gridding for various lags. The first lagged value was chosen as the discriminant metric. The number of partitions was chosen as 64 i.e. $\sqrt{N} = \sqrt{4096}$ (Tukey's criterion). As a caution it should pointed out that mutual information estimates are sensitive to the number of partitions chosen and several algorithms have been proposed in the literature.

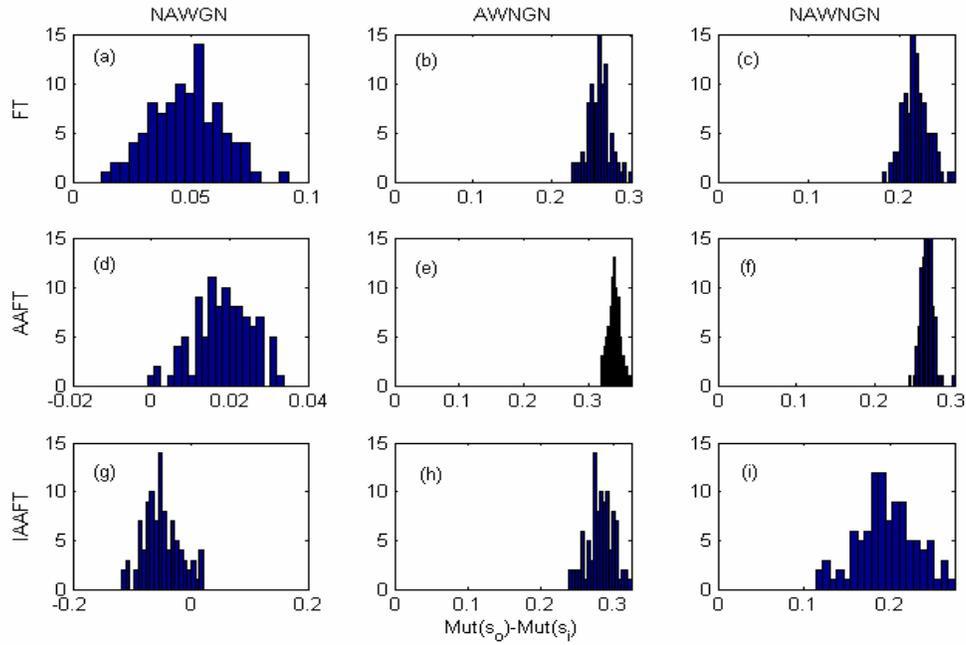

**Figure R1** Distribution of the difference in the mutual information (Mut) statistics between the given empirical sample and their FT (a, b, c), AAFT (d, e, f) and IAAFT (g, h, i) surrogate counterparts $\text{Mut}(s_o) - \text{Mut}(s_i)$, i = 1...99 for NAWGN (a, d, e) , AWNGN (b, e, h) and NAWNGN (c, f, i). The significance values obtained by parametric and non-parametric testing are enclosed under Table I (below).

**Table I** Surrogate testing using FT, AAFT and IAAFT algorithms of empirical samples generated by NAWGN, AWNGN and NAWNGN processes using first-lagged mutual information (Mut) as a discriminant statistic

| Mutual Inf. (Mut) | FT | AAFT | IAAFT |
|---|---|---|---|
| **NAWGN** | S = 0.95 | S = 2.5 | S = 1.6 |
| | p > 0.01 | p > 0.01 | p > 0.01 |
| **AWNGN** | S = 18.3 | S = 36.3 | S = 15.3 |
| | p < 0.01 | p < 0.01 | p < 0.01 |
| **NAWNGN** | S = 15.1 | S = 32.8 | S = 5.7 |
| | p < 0.01 | p < 0.01 | p < 0.01 |